\documentclass[a4paper, 12pt]{article}
 
\usepackage[left=2cm,right=2cm,top=2cm,bottom=2cm]{geometry}
\usepackage[onehalfspacing]{setspace}
\usepackage{amsmath}
\usepackage{url}
\usepackage{booktabs}

\begin{document}

\begin{titlepage}
\vspace*{-15mm}
\begin{flushright}
  TTP16-015
\end{flushright}
\vspace*{0.7cm}

\begin{center}{
\bfseries\LARGE
\texttt{E6Tensors}: A Mathematica Package for $\mathbf{E_6}$ Tensors}\\[8mm]
Thomas~Deppisch \footnote{E-mail: \texttt{thomas.deppisch@kit.edu}}\\[1mm]
\end{center}

\vspace*{0.50cm}

\centerline{\itshape
Institut f\"ur Theoretische Teilchenphysik, Karlsruhe Institute of Technology}
\centerline{\itshape
Engesserstra\ss{}e 7, D-76131 Karlsruhe, Germany
\vspace*{1.20cm}}

\begin{abstract}
We present the \texttt{Mathematica} package \texttt{E6Tensors}, a tool for explicit tensor calculations in $E_6$ gauge theories. In addition to matrix expressions for the group generators of $E_6$, it provides structure constants, various higher rank tensors and expressions for the representations $\mathbf{27}$, $\mathbf{78}$, $\mathbf{351}$ and $\mathbf{351'}$. This paper comes along with a short manual including physically relevant examples. I further give a complete list of gauge invariant, renormalisable terms for superpotentials and Lagrangians.
\end{abstract}
\end{titlepage}

\section*{Program Summary}\noindent
\textbf{Author:} Thomas Deppisch\\\noindent
\textbf{Title:} E6Tensors\\\noindent
\textbf{Licence:} GNU LGPL\\\noindent
\textbf{Programming language / external routines:} Wolfram Mathematica 8-10\\\noindent
\textbf{Operating system:} cross-platform\\\noindent
\textbf{Computer architecture:} x86, x86\_64\\\noindent
\textbf{RAM:} $>1$ GB RAM recommended\\\noindent
\textbf{Current version:} 1.0.0\\\noindent
\textbf{Web page:} \url{http://e6tensors.hepforge.org}\\\noindent
\textbf{Contact:} \url{thomas.deppisch@kit.edu} --- for bugs, possible improvements and questions

\section{Introduction}
The exceptional Lie group $E_6$ may be a suitable candidate for describing fundamental symmetries in particle physics \cite{gursey:1975ki}. In the discussion of $E_6$, most authors rely on abstract group theoretic methods. Besides, there is a paper by Kephart and Vaughn \cite{Kephart:1981gf} that describes the generators of $E_6$ in terms of its maximal subgroup $SU(3)\times SU(3)\times SU(3)$. In practice, applying these methods may be cumbersome. For writing down Lagrangians and superpotentials of such models it is useful to have explicit expressions for irreducible representations and invariants, preferably in a way suited for computer use. To our knowledge, such a tool is still missing in the literature and the package \texttt{E6Tensors} tries to fill this gap.

Our package provides such explicit expressions for the represenations $\mathbf{27}, \mathbf{78}, \mathbf{351},\mathbf{351'}$ as well as structure constants and higher rank tensors. \texttt{E6Tensors} enables the user to study Lagrangians and superpotentials including the aforementioned representations explicitly by components.

The outline of the paper is the following: In Section \ref{sec:one}, we state the transformation law for the fundamental representation $\mathbf{27}$. From that we construct the matrix expressions for the 78 group generators forming the adjoint representation $\mathbf{78}$. Then it is possible to compute the (symmetric) structure constants and other properties of the matrix generators. Since higher-order tensors can be built from the tensor products of (anti-)fundamental representations, we derive expressions for the irreducible representations $\mathbf{351}$ and $\mathbf{351'}$ in Section \ref{sec:two}. Section \ref{sec:four} then provides a small manual for the \texttt{Mathematica} package including some remarks and examples. A summary of possible gauge invariant terms in superpotentials and Lagrangians can be found in Appendix \ref{sec:three}.

\section{Matrix Expressions for the Group Generators}
\label{sec:one}

In \cite{Kephart:1981gf}, the authors give a prescription how the fundamental representation can be expressed as three $3\times 3$ matrices $L,M,N$ and how the group generators act on them. We briefly revise that prescription and  point out that we performed the calculations described there using symbolic manipulation in \texttt{Mathematica}. Throughout the paper we use Greek indices for the fundamental representation of $E_6$
\begin{equation}
  \mu,\nu,\rho,\dots = 1,\dots,27. \label{eq:indices1}
\end{equation}
For the adjoint representation we use Latin indices $k,l,m,\dots$
\begin{equation}
  k,l,m,\dots = 1,\dots,78. \label{eq:indices2}
\end{equation}
For describing $E_6$ by its maximal subgroup $SU(3)_C\times SU(3)_L \times SU(3)_R$ we also use the $SU(3)$ indices
\begin{align}
  \alpha,\beta,\gamma,\dots &= 1,2,3.\\
  a,b,c,\dots &= 1,2,3.\\
  p,q,r,\dots &= 1,2,3.\\
  A,B,C,\dots &= 1,\dots,8.
\end{align}

\subsection{Transformation Law of the Fundamental Representation}
With respect to its maximal subgroup $SU(3)_C\times SU(3)_L\times SU(3)_R$, the fundamental representation of $E_6$ can be decomposed as
\begin{equation}\label{eq:27decomposition}
  \mathbf{27} \to (\mathbf{3},\mathbf{\bar 3},\mathbf{1}) + (\mathbf{1},\mathbf{3},\mathbf{\bar 3}) + (\mathbf{\bar 3},\mathbf{1},\mathbf{3}),
\end{equation}
and its adjoint representation can be decomposed as
\begin{equation}
  \mathbf{78} \to (\mathbf{8},\mathbf{1},\mathbf{1}) + (\mathbf{1},\mathbf{8},\mathbf{1}) + (\mathbf{1},\mathbf{1},\mathbf{8}) + (\mathbf{3},\mathbf{3},\mathbf{3}) +(\mathbf{\bar 3},\mathbf{\bar 3},\mathbf{\bar 3}). \label{eq:reps}
\end{equation}

First, the generators $T^A$ of the three SU(3) subgroups corresponding to the first three representations in eq.~\eqref{eq:reps} can be represented by the eight Gell-Mann matrices $\lambda^A/2$ and their action onto the three matrices $(L,M,N)$ according to eq.~\eqref{eq:27decomposition} can be expressed by
\begin{align} \label{eq:su3onlmn}
  T^A_C\ \begin{pmatrix} L,& M,& N \end{pmatrix} &= \begin{pmatrix} \tfrac 12 \lambda^A L, & 0 ,& -\tfrac 12 N \lambda^A \end{pmatrix}, \nonumber \\
  T_L^A\ \begin{pmatrix} L,& M,& N \end{pmatrix} &= \begin{pmatrix} -\tfrac 12 L \lambda^A, & \tfrac 12 \lambda^A M, & 0 \end{pmatrix}, \\
  T_R^A\ \begin{pmatrix} L,& M,& N \end{pmatrix} &= \begin{pmatrix}0, & -\tfrac 12 M \lambda^A, & \tfrac 12 \lambda^A N\end{pmatrix}. \nonumber
\end{align}
Second, there are the generators $T^{\alpha ap}$ and $\bar T^{\alpha ap}$ that mediate shifts between the matrices $L,M,N$
\begin{align}\label{eq:shifting}
T_{\alpha a p}\ L_\beta^{\ b} &= \varepsilon_{\alpha \beta \gamma}\ \delta_a^{\ b}\ N_p^{\ \gamma}, &
\bar T^{\alpha a p} L_\beta^{\ b} &= -\varepsilon^{abc}\ \delta_\beta^{\ \alpha}\ M_c^{\ p}, \nonumber\\
T_{\alpha a p}\ M_b^{\ q} &= \varepsilon_{abc}\ \delta_p^{\ q}\ L_\alpha^{\ c}, &
\bar T^{\alpha a p} M_b^{\ q} &= -\varepsilon^{pqr}\ \delta_b^{\ a}\ N_r^{\ \alpha}, \\
T_{\alpha a p}\ N_q^{\ \beta} &= \varepsilon_{pqr}\ \delta_\alpha^{\ \beta}\ M_\alpha^{\ r}, &
\bar T^{\alpha a p} N_q^{\ \beta} &= -\varepsilon^{\alpha\beta\gamma}\ \delta_q^{\ p}\ L_\gamma^{\ a}. \nonumber
\end{align}
$\delta_a^{b}$ is the Kronecker symbol and $\varepsilon^{abc}$ the Levi-Civita symbol with $\varepsilon_{123} = \varepsilon^{123}=1$. With these set of generators an infinitesimal, unitary $E_6$ transformation reads
\begin{equation}\label{eq:infinite_transformation1}
U(u,v,w,x,y) = \mathbf{1} + \mathrm{i}\, u_A T^A_C + \mathrm{i}\, v_A T^A_L + \mathrm{i}\, w_A T^A_R + \mathrm{i}\, x_{\alpha ap} T^{\alpha ap} + \mathrm{i}\, y_{\alpha ap} \bar T^{\alpha ap} + \dots
\end{equation}
In total, $u_A,v_A,w_A,x_{\alpha ap},y_{\alpha ap}$ are 78 parameters.

\subsection{Explicit Matrix Expressions for the Group Generators}
We now aim at writing the transformation in eq. \eqref{eq:infinite_transformation1} with one set of 78 matrices $_k T$ and parameters $\varepsilon^k$ that act on a 27-dimensional vector $\psi$
\begin{equation}\label{eq:infinite_transformation2}
U(\varepsilon)\ \psi =  \left( \mathbf{1}+\mathrm{i}\ \varepsilon^k\, _k T + \dots \right) \psi.
\end{equation}
For that purpose, we rearrange the transformation parameters and the matrices $L,M,N$ into column vectors in the following way:
\begin{align}\label{eq:flatten}
  \psi &= (L_{11},L_{12},...,M_{11},...,N_{11},...,N_{33})^T \nonumber\\
  \varepsilon &= (u^1,u^2,...,v^1,...,w^1,...,x^{111},...,y^{111},...,y^{333})^T.
\end{align}
By comparing coefficients in \eqref{eq:infinite_transformation1} and \eqref{eq:infinite_transformation2}, the $27\times 27$ matrices $_kT$ can be constructed. To adjust the normalisation and obtain Hermiticity, we perform the following change of basis
\begin{align}
  _k\tilde T &= \frac 12 \left(\, _k T +\, _{k+27} T \right), &24<k<52 \nonumber\\
  _k\tilde T &= \frac{\mathrm{i}}{2} \left(\, _k T -\, _{k+27} T \right), &51<k<78.
\end{align}
The group generators are included in the \texttt{Mathematica} package as an $78\times 27\times 27$ dimensional array called \texttt{E6gen}. They obey
\begin{align}
  \operatorname{tr} \left(\, _k T \, _l T \right) &= 3\ \delta_{kl}\\
  \sum\limits_{k=1}^{78} \, _kT \, _kT &= \tfrac{26}{3}\ \mathbf{1}_{27}. \label{eq:casimir}
\end{align}
This sets the Dynkin index and the quadratic Casimir invariant to
\begin{equation}
  C(27) = 3 \qquad \mathrm{and} \qquad C_2(27) = \tfrac{26}{3},
\end{equation}
satisfying the well-known identity
\begin{align}
  C_2(R) &= \frac{\operatorname{dim}(G)}{\operatorname{dim}(R)}\ C(R),
\end{align}
for a representation $R$ and the adjoint representation $G$. In addition to this consistency check, $C(27)$ and $C_2(27)$ match the same values Kephart and Vaughn state in their paper \cite{Kephart:1981gf}.

By construction, the generators are ordered in the following way
\begin{align}
  _1T \dots \, _8T &:\qquad SU(3)_C,\\
  _{9}T \dots\, _{16} T &:\qquad SU(3)_L,\\
  _{17}T \dots\, _{24} T &:\qquad SU(3)_R.
\end{align}
Therefore, the diagonal generators representing the Cartan subalgebra are
\begin{equation}
  _3T,\,_8 T,\, _{11} T,\, _{16} T,\, _{19} T,\, _{24} T.
\end{equation}
The generators $_{25}T$ to $_{78}T$ are the shifting operators defined in eq. \eqref{eq:shifting}. The structure constants $f_{klm}$ of a Lie Algebra are defined by
\begin{equation}
  [\, _kT,\, _lT] = \mathrm{i}f_{klm} \, _m T.
\end{equation}
Applying the normalisation condition gives
\begin{equation}
  f_{klm} = -\dfrac{\mathrm{i}}{C(27)} \operatorname{tr} \left( [\, _k T,\, _l T]\ _m T \right).
\end{equation}
In the \texttt{Mathematica} package they are encoded in the array \texttt{E6f}. As a cross-check we calculated the normalisation to be
\begin{equation}
 f_{kmn}f_{lmn} = 12\ \delta_{kl}
\end{equation}
which also matches the value in \cite{Kephart:1981gf}. Since the structure constants are the generators of the adjoint representation, its quadratic Casimir and Dynkin index are
\begin{equation}
  C(G)=C_2(G)=12.
\end{equation}
The symmetric structure constants $C_{klm}$ are defined by
\begin{equation} \label{eq:anti-commutator}
 \{\, _kT,\, _lT\}= \, _kT\, _l T + \, _l T \, _k T = \mathrm{i}\ C_{klm} \, _m T.
\end{equation}
This can be rewritten as
\begin{equation} \label{eq:symmetric_structure_constants}
  C_{klm} = -\dfrac{\mathrm{i}}{C(27)} \operatorname{tr} \left( \{\, _kT,\, _l T\}\ _mT \right ).
\end{equation}
Explicit computation then yields
\begin{equation}
  C_{klm} = 0 \quad \forall\ k,l,m =1,\dots,78.
\end{equation}
Hence, $E_6$ GUT models are in general free of chiral anomalies.

\section{Higher Rank Tensors}
\label{sec:two}

The transformation laws for fundamental \textbf{27}, anti-fundamental $\overline{\mathbf{27}}$ and adjoint representation \textbf{78} are as follows
\begin{align}
  \psi_\mu &\to \psi_\mu + \mathrm{i}\,\varepsilon^k\ (_k T)_\mu^{\,\nu} \psi_\nu , \label{eq:psi}\\
  \bar \psi^\mu &\to \bar\psi ^\mu- \mathrm{i}\,\varepsilon^k\ (_k T)_\nu^{\,\mu} \bar\psi^\nu, \label{eq:psibar}\\
  \phi_l &\to \phi_l + \varepsilon^k\ f_{klm} \phi_m,
\end{align}
with Greek indices running from 1 to 27 and Latin indices running from 1 to 78, cf. eq.~\eqref{eq:indices1} and eq.~\eqref{eq:indices2}.

\subsection{Higher Dimensional Representations}

The representations $\mathbf{351}$, $\mathbf{351'}$ and $\mathbf{650}$ are included in the tensor products \cite{Slansky:1981yr}\footnote{The notation for $\mathbf{351}$ and $\mathbf{351'}$ differs in the literature. In our notation, $\mathbf{351}$ is symmetric, whereas $\mathbf{351'}$ is anti-symmetric.}
\begin{align}
  \overline{\mathbf{27}} \otimes \overline{\mathbf{27}} &= \mathbf{27} \oplus \mathbf{351} \oplus \mathbf{351'},\\
  \overline{\mathbf{27}} \otimes \mathbf{27} &= \mathbf{1} \oplus \mathbf{78} \oplus \mathbf{650}.
\end{align}
Therefore, they can be represented as rank-two tensors. Their transformation properties are implicitly given by \eqref{eq:psi} and \eqref{eq:psibar}. For $\mathbf{351}$ and $\mathbf{351'}$, we labelled the entries for that tensors $\chi_1,\dots,\chi_{351}$ and choose them in a way that
\begin{equation}
  \bar X^{\mu\nu} X_{\mu\nu} = \bar\chi^1 \chi_1 + \dots + \bar\chi^{351} \chi_{351}, \qquad \mu,\nu=1,\dots 27,
\end{equation}
ensuring a canonical normalisation of the kinetic term. A comment on this normalisation is given in Appendix \ref{app:b}.

$\mathbf{351}$ can be represented by a second rank tensor $A_{\mu\nu}$ antisymmetric in its indices. In the \texttt{Mathematica} package it is included as an $27\times 27$ dimensional array called \texttt{E6A}.

$\mathbf{351'}$ is also a rank two tensor $S_{\mu\nu}$ but symmetric in its indices and additionally satisfying
\begin{equation}
  d^{\mu\nu\lambda}  S_{\mu\nu} = 0, \qquad \forall\ \lambda=1,\dots,27,
\end{equation}
with $d^{\mu\nu\lambda}$ defined below in eq. \eqref{eq:cubic}. $\mathbf{351'}$ is named \texttt{E6S} in the package.

\textbf{650} has a fundamental and an anti-fundamental index with vanishing trace $\psi_\mu^{\ \mu} =0$ and 
\begin{equation}
  _kT_\nu^{\,\mu}\, \psi_\mu^{\,\nu} = 0.
\end{equation}
It is not (yet) included in the package due to its memory usage. 

\subsection{Invariants for the Fundamental Representation}

The Kronecker symbol $\delta^\mu_\nu$ is the most simple way to define a quadratic invariant of a fundamental and an anti-fundamental repesentation
\begin{equation}
  \delta^\mu_\nu\, \bar\psi^\nu \psi_\mu.
\end{equation}

A cubic invariant can be defined in the following way \cite{Kephart:1981gf}
\begin{equation} \label{eq:cubic}
  d^{\mu\nu\lambda}\, \psi_\mu\psi_\nu\psi_\lambda = \det{(L + M + N)} - \operatorname{tr}(LMN).
\end{equation}
The entries of the tensor $d^{\mu\nu\lambda}$ can be obtained by comparing the coefficients of the field components on each side of the equation. It is provided as \texttt{E6d} in the \texttt{Mathematica} package. Together with $d_{\mu\nu\lambda}$, defined by
\begin{equation}
  d_{\mu\nu\lambda}\, \bar\psi^\mu \bar\psi^\nu\bar\psi^\lambda = \det{(L^\dagger + M^\dagger + N^\dagger)} - \operatorname{tr}(N^\dagger M^\dagger L^\dagger),
\end{equation}
it is normalised to
\begin{equation}
  d_{\mu\nu\lambda}d^{\mu\nu\sigma} = 10\, \delta_\lambda^\sigma.
\end{equation}

Additionally there are some compound tensors that are used to construct the invariants in Appendix~\ref{sec:three}. A tensor  with four indices is
\begin{equation}
  D_{\mu\nu}^{\sigma\tau}  = d_{\mu\nu\lambda}d^{\lambda\sigma\tau}.
\end{equation}
In the package it is called \texttt{E6D}.

There is also a compound tensor carrying five indices:
\begin{align}
  D_\lambda^{\mu\nu,\sigma\tau} &= (d^{\mu\sigma\xi} d^{\nu\tau\eta}-d^{\mu\tau\xi} d^{\nu\sigma\eta})d_{\xi\eta\lambda},\\
  D^\lambda_{\mu\nu,\sigma\tau} &= (d_{\mu\sigma\xi} d_{\nu\tau\eta}-d_{\mu\tau\xi} d_{\nu\sigma\eta})d^{\xi\eta\lambda}.
\end{align}
They are not included in the package to avoid excessive memory usage.

\subsection{Invariants for the Adjoint Representation}

The normalisation condition for the generators can be used to define a quadratic invariant for the adjoint representation $\phi$.
\begin{equation}
  \delta_{kl}\ \phi_k\phi_l = \tfrac 13 \operatorname{tr}(_kT_lT)\phi_k\phi_l
\end{equation}
With the structure constants one can form an invariant of three different adjoint representations $\phi,\phi',\phi''$:
\begin{equation}
  f_{klm}\ \phi_k\phi'_l\phi''_m.
\end{equation}
The completely symmetric tensor
\begin{equation}
  \chi^5_{klmn} = \delta_{kl}\delta_{mn} +\delta_{kl}\delta_{mn} +\delta_{kl}\delta_{mn}
\end{equation}
in the package is called \texttt{E6chi} and can be used to form an quartic invariant $\chi^5_{klmn}\phi'_k\phi''_l\phi'''_m\phi''''_n$.

\subsection{Mixed Invariants}

The generators $(_kT)_\mu^{\,\nu}$ form a tensor $_kT_\mu^{\,\nu}$ with an adjoint, a fundamental and an anti-fundamental index.

Further, there is a tensor that can be constructed from the anti-commutator and Kronecker symbols
\begin{equation}
  _{kl}H_\mu^{\,\nu} = \left\{T_k,T_l\right\}_\mu^{\,\nu} - \tfrac 29 \,\delta_{kl} \delta_\mu^\nu,
\end{equation}
called \texttt{E6H} in the package.

Contracting $_kT_\mu^{\,\nu}$ with $d_{\nu\rho\lambda}$ gives the tensor
\begin{equation}
  _kA_{\mu\nu,\lambda} = \,_kT_\mu^{\ \sigma} d_{\sigma\nu\lambda} - \,_kT_\nu^{\ \sigma} d_{\mu\sigma\lambda}
\end{equation}
which is antisymmetric w.r.t. $\mu\leftrightarrow\nu$ and the tensor
\begin{equation}
    _kS_{\mu\nu,\lambda} = -\,_kT_\lambda^{\ \sigma} d_{\mu\nu\sigma}
\end{equation}
which is symmetric w.r.t. $\mu\leftrightarrow\nu$. They are called \texttt{E6kA} and \texttt{E6kS}, respectively.

\section{\texttt{E6Tensors.m} - A short Manual}
\label{sec:four}

\subsection{Download and Installation}
\texttt{E6Tensors} can be downloaded from \url{http://e6tensors.hepforge.org}. On that page, there are also some instructions on how to install it. Currently, there are two versions at the download section: \texttt{e6tensors\_full-1.0.0.tar.gz} and \texttt{e6tensors\_small-1.0.0.tar.gz}.

\texttt{e6tensors\_small-1.0.0.tar.gz} contains the following files: \texttt{install.sh} calls the command line version of \texttt{Mathematica}\footnote{We assume it to be called \texttt{math}. Change that if it has another name on your system.} and runs \texttt{create\_E6Tensors.m}. This script uses \texttt{E6gen.m} an \texttt{E6d.m} to create the higher dimensional tensors and saves them as arrays to \texttt{E6Tensors.m}. \texttt{examples.nb} shows some well-documented examples how \texttt{E6Tensors.m} can be used. Note that \texttt{E6Tensors.m} has a size of roughly 150 MB. Therefore make sure to provide enough RAM for loading it. Running \texttt{create\_E6Tensors.m} also may need some time. On a quad-core i5 machine this took about half an hour working on four subkernels. To use parallelisation, change \texttt{LaunchKernels[1]} in \texttt{create\_E6Tensors.m} to the appropriate value.

For most users, we recommend to download \texttt{e6tensors\_full-1.0.0.tar.gz}. After extracting the tarball, it is ready to use and contains all files of \texttt{e6tensors\_small-1.0.0.tar.gz} including \texttt{E6Tensors.m}.

You probably will not need all tensors in a single project. In that case you can comment out the unnecessary tensors in \texttt{create\_E6Tensors.m} and run it to get your own customised file \texttt{E6Tensors.m}.

\subsection{Structure of the Package}
\texttt{E6Tensors.m} is a simple text file. It contains the definitions of all tensors as nested lists. In this way, it is very flexible to use: You can write your own functions and procedures that fit the problem you want to solve. As an example, the Pauli matrices would look like
\begin{verbatim}
{{{0,1},{1,0}},{{0,-I},{I,0}},{{1,0},{0,-1}}}.
\end{verbatim}
The generators \texttt{E6gen} have exactly the same structure. Hence,
\begin{verbatim}
E6gen[[k,mu,nu]]
\end{verbatim}
is the element in the $\mu^{\mathrm{th}}$ row and the $\nu^{\mathrm{th}}$ column of the $k^{\mathrm{th}}$ generator. All tensors are listed in Table~\ref{tab:tensors}. There, you can also find their symbolic name, the indices they carry and a short explanation.

For instance, \texttt{E6gen} has indices $78,\overline{27},{27}$ refering to the gauge index $k=1,\dots,78$, the row index $\mu=1,\dots,27$ and the column index $\nu=1,\dots,27$. The order follows the convention of the \texttt{Part[]} function in \texttt{Mathematica}. Keep in mind that \texttt{Mathematica} does not make any difference between row and column vectors.
\begin{table}
  \centering
  \begin{tabular}{llll}
    \toprule[.5pt]
    Name & Name & Indices & Comment \\
    in \cite{Kephart:1981gf} & in \texttt{E6Tensors} & & \\
    \midrule
    $_kT_\mu^{\, \nu}$ & \texttt{E6gen} & $78,\overline{27},27$ & adjoint representation $\mathbf{78}$ \\
    &&& $\operatorname{tr} \left( _kT\, _l T \right) = 3\, \delta_{kl}$\\[2mm]
    $A_{\mu\nu}$ & \texttt{E6A} & 27,27 & antisymmetric $\mathbf{351}$ \\
    &&&$27\times 27$ matrix with entries labelled $\chi_1\dots\chi_{351}$\\[2mm]
    $S_{\mu\nu}$ & \texttt{E6S} & 27,27 & symmetric $\mathbf{351'}$\\
    &&& $d^{\mu\nu\sigma}S_{\nu\sigma}=0$\\
    &&& $27\times 27$ matrix with entries labelled $\chi_1\dots\chi_{351}$\\[2mm]
    $d_{\mu\nu\lambda}$ & \texttt{E6d} & 27,27,27 & fully symmetric invariant\\ 
    &&& $d_{\mu\nu\lambda} d^{\mu\nu\sigma} = 10\, \delta_\lambda^\sigma$ \\[2mm]
    $f_{klm}$ & \texttt{E6f} & 78,78,78 & structure constants: $[\, _kT,\, _lT]= \mathrm{i}f_{klm}\, _m T$\\
    &&&$f_{kmn}f_{lmn} = 12\, \delta_{kl}$\\[2mm]
    $D^{\sigma\tau}_{\mu\nu}$ & \texttt{E6D} & $\overline{27},\overline{27},27,27$ & $D^{\sigma\tau}_{\mu\nu} = \delta^\sigma_\mu \delta^\tau_\nu + \delta^\tau_\nu \delta^\sigma_\mu$ \\[2mm]
    $\chi^5_{klmn}$ & \texttt{E6chi} & 78,78,78,78 & $\chi^5_{klmn} = \delta_{kl}\delta_{mn} + \delta_{km}\delta_{ln} + \delta_{kn}\delta_{lm}$\\[2mm]
    $_{kl}H_\mu^{\,\nu}$ & \texttt{E6H} & $27,\overline{27},78,78$ & $_{kl}H_\mu^{\,\nu} = \{ T_k, T_l\}_\mu^{\;\nu} - \tfrac 29\, \delta_{kl} \delta_\mu^{\;\nu}$\\[2mm]
    $_k A_{\mu\nu,\lambda}$ & \texttt{E6kA} & 78,27,27,27 & antisymmetric in $\mu,\nu$\\
    &&& $_kA_{\mu\nu,\lambda} =\, _kT_\mu^{\;\sigma}d_{\sigma\nu\lambda} -\, _kT_\nu^{\;\sigma}d_{\mu\sigma\lambda}$\\[2mm]
    $_k S_{\mu\nu,\lambda}$ & \texttt{E6kS} & 78,27,27,27 & symmetric in $\mu,\nu$\\
    &&& $_kS_{\mu\nu,\lambda} =- _kT_\lambda^{\;\sigma}d_{\mu\nu\sigma}$\\[2mm]
    \bottomrule[.5pt]
  \end{tabular}
  \caption{Overview of Tensors in \texttt{E6Tensors.m}.}
  \label{tab:tensors}
\end{table}

\subsection{Known Issues}
It is not recommended to open \texttt{E6Tensors.m} via the graphical frontend of \texttt{Mathematica}. To load it use
\begin{verbatim}
Get["##/E6Tensors.m"]
\end{verbatim}
instead, where \texttt{\#\#} refers to the correct path.

\subsection{Examples}
In the download version, there is a notebook \texttt{examples.nb }hat contains some possible applications of the package. It starts with loading \texttt{E6gen.m} and \texttt{E6d.m} which are sufficient for the first examples.

We identify the Standard Model generators among the $E_6$ generators: By construction, we can choose the gluons to be the first eight generators. The generators of $SU(2)_L$ can then be defined as
\begin{equation}
  T^{9}, T^{10}, T^{11},
\end{equation}
and hypercharge as
\begin{equation}
  Y = \sqrt{\frac 35} \left(-\sqrt{\frac 13} T^{16} - T^{19}  - \sqrt{\frac 13} T^{24} \right).
\end{equation}
There are two additional U(1) charges, which we can define by
\begin{align}
  Y' &= \sqrt{\frac{1}{40}} \left( -2\sqrt{3} T^{16} - T^{19} + 3 \sqrt{3} T^{24}\right),\\
  Y'' &= \sqrt{\frac{1}{40}} \left( -2\sqrt{3} T^{16} +4 T^{19} -2 \sqrt{3} T^{24}\right).
\end{align}
In this basis, there is a singlet in the fundamental representation for $Y'$ and $Y''$ each.

The generators for $SU(2)_R$ can be defined as
\begin{equation}
  T^{17},T^{18},T^{19},
\end{equation}
and $B-L$ as
\begin{equation}
  B-L = -\sqrt{\frac 12}\left(T^{16} + T^{24}\right) =\sqrt{\frac 52} Y + \sqrt{\frac 32}T^{19}.
\end{equation}

As a first check, we can write the fundamental representation as a list of field names and show their quantum numbers in a table. We also check for the GUT normalisation of the $U(1)$ charges and the correct commutation relation for $SU(2)_L$, $SU(2)_R$ and $Y$.

In a next step, we use $d^{\mu\nu\lambda}$ to write down the trilinear coupling in the superpotential
\begin{equation}
   \mathcal{W} = \frac{\lambda}{6}\ d^{\mu\nu\rho} \psi_\mu\psi_\nu\psi_\rho.
\end{equation}
For instruction, we once write out the explicit sum over all indices and once use the \texttt{Dot[]} operator. In many cases, the latter one will be the faster way to do it. The same holds for functions like \texttt{TensorContract[]}.

For the next examples \texttt{E6Tenors.m} must be located in the same directory. We first test the tensors for the higher dimensional representations $\mathbf{351}$ and $\mathbf{351'}$, i.e. their defining properties
\begin{equation}
  d^{\mu\nu\lambda} S_{\mu\nu} = 0, \qquad S_{\mu\nu} = S_{\nu\mu},
\end{equation}
and
\begin{equation}
  A_{\mu\nu} = - A_{\nu\mu}.
\end{equation}
The normalisation of the kinetic terms gives the wanted result.

The couplings to matter fields can be described by $\mathcal{W}= \bar S^{\mu\nu} \psi_\mu \psi_\nu $ and $\mathcal{W}= \bar A^{\mu\nu} \psi_\mu \psi_\nu$. Now, we can read off the fields that couple to down quarks. Since $A^{\mu\nu}$ is anti-symmetric, it does not couple fields of the same representation (e.g. the same flavour) to each other.

For a more advanced example, we discuss possible vacuum expectation values (VEVs) for $\mathbf{351'}$ ($S^{\mu\nu}$) and $\mathbf{351}$ ($A^{\mu\nu}$). Since they are contained in the tensor product $\mathbf{27}\otimes \mathbf{27}$, we can write an infinitesimal $E_6$ transformation as
\begin{equation}
  S^{\mu\nu} \to \left(\delta^\mu_\rho + \mathrm{i}\, \alpha_k\,_k T_\rho^{\,\mu} \right)
\left(\delta^\nu_\sigma + \mathrm{i}\, \alpha_k\,_k T_\sigma^{\,\nu} \right) S^{\rho\sigma}
= S^{\mu\nu} + \mathrm{i}\, \alpha_k \left(_kT_\nu^{\,\sigma} S_{\mu\sigma}+ \,_kT_\mu^{\,\rho} S_{\rho\nu} \right). \label{eq:translaw}
\end{equation}
$\alpha_k$ is a set of parameters. For a VEV $s^{\mu\nu}=\langle S^{\mu\nu}\rangle$ that transforms trivially under a set of generators $\{_kT\}$, the last term in \eqref{eq:translaw} must vanish. Using the permutation symmetry of $S^{\mu\nu}$, this can be written as a matrix equation
\begin{equation}
  (_kT\cdot s)^T + \,_kT\cdot s = 0.
\end{equation}
$A^{\mu\nu}$ is antisymmetric, therefore the condition reads
\begin{equation}
    (_kT\cdot a)^T - \,_kT\cdot a = 0.
\end{equation}
The conditions are implemented in \texttt{example.nb}. As set of generators we use the gluons and
\begin{equation}
  Q = I_L^3 + \sqrt{\frac 53} Y.
\end{equation}
This ensures that the vacuum does not carry electric charge or colour. We then calculate the resulting mass terms that are generated by this VEV.

\section{Acknowledgements}
The author likes to thank Martin Spinrath for helpful discussions and commments on this paper as well as Julia Gehrlein, Jakob Schwichtenberg and Georg Winner for testing the package and Ulrich Nierste for proofreading. This work was supported by \emph{Studienstiftung des deutschen Volkes} and the DFG-funded Research Training Group \emph{GRK 1694 - Elementarteilchenphysik bei h\"ochster Energie und h\"ochster Pr\"azision}.

\begin{appendix}
\section{Renormalisable Potentials}
\label{sec:three}

For completeness, we write down all possible renormalisable and gauge invariant terms that may occur in superpotentials and Lagrangians. They can also be found in \cite{Kephart:1981gf}.

\subsubsection*{Mass Terms}
Possible mass terms for the various representations are:
\begin{align*}
  \overline{\mathbf{27}}\cdot\mathbf{27} \qquad&: \qquad \bar\psi^\mu\psi_\mu\\
  \mathbf{78}^2 \qquad&: \qquad \operatorname{tr}(_kT_lT) \phi_k \phi'_l= 3\, \phi_k\phi'_k \\
  \overline{\mathbf{351}}\cdot\mathbf{351} \qquad&:\qquad \bar A^{\mu\nu}A_{\mu\nu}\\
  \overline{\mathbf{351'}}\cdot\mathbf{351'} \qquad&:\qquad \bar S^{\mu\nu}S_{\mu\nu}
\end{align*}

\subsubsection*{Cubic Terms}
\begin{align*}
  \mathbf{27}^3\qquad &:\qquad d^{\mu\nu\lambda}\psi_\mu\psi_\nu\psi_\lambda\\
  \mathbf{351'}^3\qquad &:\qquad d^{\mu\nu\lambda}d^{\sigma\tau\rho}S_{\mu\sigma}S_{\nu\tau}S_{\lambda\rho}\\
  \mathbf{27}^2\cdot\overline{\mathbf{351'}}\qquad &:\qquad \psi_\mu\psi_\nu \bar S^{\mu\nu}\\
  \mathbf{351}^2\overline{\mathbf{27}}\qquad &:\qquad D_\lambda^{\mu\nu,\sigma\tau} A_{\mu\nu}A_{\sigma\tau} \bar\psi^\lambda\\
  \mathbf{27}\cdot\mathbf{351}\cdot\mathbf{78}\qquad &:\qquad _kT_\mu^{\,\tau}d^{\mu\sigma\lambda} \psi_\lambda A_{\sigma\tau} \phi_k\\
  \mathbf{78}^3 \qquad &: \qquad f_{klm}\phi_k \phi'_l \phi''_m \\
  \mathbf{27}\cdot\overline{\mathbf{27}}\cdot\mathbf{78}\qquad &:\qquad _kT_\mu^{\,\nu} \bar\psi^\mu \psi_\nu \phi_k\\
  \mathbf{351}\cdot\overline{\mathbf{351}}\cdot\mathbf{78}\qquad &:\qquad _kT_\mu^{\,\nu} \bar A^{\mu\lambda} A_{\nu\lambda} \phi_k\\
  \mathbf{351'}\cdot \mathbf{351'}\cdot \mathbf{78} \qquad &: \qquad _kT_\mu^{\,\nu} \bar S^{\mu\lambda}S_{\nu\lambda} \phi_k
\end{align*}

\subsubsection*{Quartic Terms}
For the fundamental and adjoint representations, there are
\begin{align*}
  \mathbf{27}^2\cdot \overline{\mathbf{27}}^2&:&\bar\psi^\mu\psi_\mu \bar\psi^\nu\psi_\nu \\
  &&D^{\mu\nu}_{\sigma\tau}\bar\psi^\sigma\bar\psi^\tau \psi_\mu \psi_\nu\\
  \mathbf{78}^4&:&(\phi_k\phi_k)^2\\
  && (\phi_l\phi_l)^2(\phi'_k\phi'_k)^2\\
  && (\phi_l\phi'_l)^2(\phi_k\phi_k')^2\\
  && \phi_k\phi_l\phi_m\phi_n \operatorname{tr}\bigl( \{_kT,_lT\}\{_mT,_nT\}\bigr)\\
  && \chi^5_{klmn}\phi_k\phi_l\phi_m\phi_n\\
  \mathbf{27}\cdot \overline{\mathbf{27}} \cdot \mathbf{78}^2 &:& \bar\psi^\mu\psi_\mu \phi_k \phi_k\\
  && (_{kl}H_\mu^{\,\nu}) \bar\psi^\mu \psi_\nu \phi_k\phi_l
\end{align*}
Including $\mathbf{351}$ and $\mathbf{351'}$ gives
\begin{align*}
  \mathbf{351}^2\overline{\mathbf{351}}^2 &:& (A_{\mu\nu}\bar A^{\mu\nu})^2\\
  && A_{\mu\nu}\bar A^{\nu\sigma}A_{\sigma\tau}\bar A^{\tau\mu}\\
  && d^{\mu\nu\lambda}d_{\xi\eta\lambda} A_{\mu\sigma}A_{\nu\tau}\bar A^{\xi\sigma} \bar A^{\eta\tau}\\
  && d^{\mu\nu\alpha}d^{\sigma\tau\beta}d_{\xi\eta\alpha}d_{\lambda\rho\beta} A_{\mu\sigma} \bar A_{\nu\tau} A^{\xi\lambda} \bar A^{\eta\rho}\\
  && d_{\mu\nu\alpha} d^{\sigma\beta\gamma} d_{\xi\eta\beta} d_{\lambda\alpha\gamma} A_{\mu\sigma} A_{\nu\tau} \bar A_\xi^\lambda \bar A^{\eta\tau} \\
  && d^{\mu\nu\alpha} d^{\sigma\tau\beta} d_{\alpha\beta\gamma} d^{\gamma\xi\chi} d_{\xi\eta\zeta} d_{\lambda\rho\chi} A_{\mu\sigma} A_{\nu\tau} \bar A^{\xi\lambda} \bar A^{\eta\rho}\\
  \mathbf{351'}^2 \overline{\mathbf{351'}}^2  &:& (A_{\mu\nu}\bar A^{\mu\nu})^2\\
  && S_{\mu\nu}\bar S^{\nu\sigma}S_{\sigma\tau}\bar S^{\tau\mu}\\
  && d^{\mu\nu\lambda}d_{\xi\eta\lambda} S_{\mu\sigma}S_{\nu\tau}\bar S^{\xi\sigma} \bar S^{\eta\tau}\\
  && d^{\mu\nu\alpha}d^{\sigma\tau\beta}d_{\xi\eta\alpha}d_{\lambda\rho\beta} S_{\mu\sigma} \bar S_{\nu\tau} S^{\xi\lambda} \bar S^{\eta\rho}\\
  \mathbf{351}\cdot \overline{\mathbf{351}}\cdot\mathbf{78}^2 &:& \bar A^{\mu\nu}A_{\mu\nu} \phi_k\phi_k\\
  && (_{kl}H_\mu^{\,\nu}) \bar A^{\mu\lambda}A_{\nu\lambda} \phi_k\phi_l\\
  && (_{kl}H_\mu^{\,\nu}) d^{\mu\sigma\alpha} d_{\nu\tau\alpha} \bar A^{\tau\lambda} A_{\sigma\lambda} \phi_k\phi_l\\
  && (_kT_\mu^{\,\sigma}) d^{\mu\lambda\alpha} (_lT_\tau^{\,\nu}) d_{\nu\rho\alpha} \bar A^{\rho\tau} A_{\alpha\lambda} \phi_k \phi_l\\
  \mathbf{351'}\cdot \overline{\mathbf{351'}}\cdot\mathbf{78}^2 &:& \bar S^{\mu\nu}S_{\mu\nu} \phi_k\phi_k\\
  && (_{kl}H_\mu^{\,\nu}) \bar S^{\mu\lambda} S_{\nu\lambda} \phi_k\phi_l\\
  && (_{kl}H_\mu^{\,\nu}) d^{\mu\sigma\alpha} d_{\nu\tau\alpha} \bar S^{\tau\lambda} S_{\sigma\lambda} \phi_k\phi_l\\
\mathbf{27}\cdot \overline{\mathbf{27}}\cdot \mathbf{351} \cdot \overline{\mathbf{351}} &:& \bar\psi^\mu\psi_\mu \bar A^{\sigma\tau} A_{\sigma\tau}\\
  && \bar\psi^\mu\psi_\nu \bar A^{\nu\tau}A_{\mu\tau}\\
  && d_{\mu\nu\lambda} d^{\xi\eta\lambda} \bar \psi^\mu \psi_\xi \bar A^{\nu\tau} A_{\eta\tau}\\
\mathbf{27}\cdot \overline{\mathbf{27}}\cdot \mathbf{351'} \cdot \overline{\mathbf{351'}} &:& \bar\psi^\mu\psi_\mu \bar S^{\sigma\tau} S_{\sigma\tau}\\
  && \bar\psi^\mu\psi_\nu \bar S^{\nu\tau}S_{\mu\tau}\\
\mathbf{27}^2 \mathbf{351}^2 &:& d^{\mu\sigma\xi}d^{\nu\tau\eta} \psi_\mu \psi_\nu A_{\sigma\tau} A_{\xi\eta}\\
  && d^{\mu\sigma\alpha} d^{\nu\xi\beta} d_{\alpha\beta\gamma}d^{\gamma\tau\eta} \psi_\mu \psi_\nu A_{\sigma\tau} A_{\xi\eta} \\
  \mathbf{27}^2 \mathbf{351'}^2 &:& d^{\mu\sigma\xi}d^{\nu\tau\eta} \psi_\mu \psi_\nu S_{\sigma\tau} S_{\xi\eta}\\
  \mathbf{351}^3 \mathbf{78} &:& (_kT_\rho^\eta) d^{\mu\sigma\alpha} d^{\nu\tau\beta} d_{\alpha\beta\gamma} d^{\gamma\xi\rho} A_{\mu\nu} A_{\sigma\tau} A_{\xi\eta} \phi_k
\end{align*}

\section{On the Normalisation of $\mathbf{351'}$}
\label{app:b}

The symmetric tensor $\mathbf{351'}$ ($S_{\mu\nu}$) is defined by
\begin{equation}
  S_{\nu\mu} = S_{\mu\nu} \qquad \mathrm{and}\qquad d^{\mu\nu\lambda}S_{\mu\nu}=0\quad \forall\ \lambda=1,\dots,27.
\end{equation}
The first condition is easy to construct: We label the off-diagonal entries $\phi_1,\dots,\phi_{351}$ and the diagonal ones $\phi_{352},\dots,\phi_{378}$. The second condition then eliminates 27 entries. For $\lambda=1$, it reads
\begin{equation} \label{eq:condition}
  \phi_{122}+ \phi_{207}+\phi_{226}+\phi_{244}-\phi_{102}= 0.
\end{equation}
It is now tempting to solve e.g. for $\phi_{102}$ and subsitute that in $S_{\mu\nu}$. But then the kinetic term $\partial^\alpha S^{\mu\nu}\partial_\alpha S_{\mu\nu}$ is not canonically normalised.\footnote{$\alpha$ is a space-time index in this case.} Another solution is to introduce new field names $\psi_1,\psi_2,\psi_3,\psi_4$ with
\begin{align}
  \phi_{102} &= a (\psi_1+\psi_2+\psi_3+\psi_4)\\
  \phi_{122} &= \psi_1-b(\psi_1+\psi_2+\psi_3+\psi_4)\\
  \phi_{207} &= \psi_2-b(\psi_1+\psi_2+\psi_3+\psi_4)\\
  \phi_{226} &= \psi_3-b(\psi_1+\psi_2+\psi_3+\psi_4)\\
  \phi_{244} &= \psi_4-b(\psi_1+\psi_2+\psi_3+\psi_4)
\end{align}
For $a=1-4b$ and $b=(5+\sqrt{5})/20,$ the defining condition is fulfilled and the kinetic term for $S^{\mu\nu}$ takes the form
\begin{equation}
  \partial_\alpha \bar S^{\mu\nu}  \partial^\alpha S_{\mu\nu} = \dots 
+ \partial_\alpha\bar \psi_1 \partial^{\alpha}\psi_1 
+ \partial_\alpha\bar \psi_2 \partial^{\alpha}\psi_2 
+ \partial_\alpha\bar \psi_3 \partial^{\alpha}\psi_3 
+ \partial_\alpha\bar \psi_4 \partial^{\alpha}\psi_4 
+ \dots
\end{equation}
The same procedure also works for all other values of $\lambda$. It is important, that the component with the relative minus sign ($\phi_{102}$ in eq.~\eqref{eq:condition}) is replaced by the expression with $a$ in it. This procedure is implemented in \texttt{create\_E6Tensors.m} and used to construct \texttt{E6S}.

\end{appendix}

\bibliographystyle{plain}
\bibliography{literature}
\end{document}